\documentclass[onecolumn]{emulateapj}

\usepackage{graphicx}
\usepackage{epsfig}

\shorttitle{Pulse Profiles with NICER}
\shortauthors{\"Ozel et al.}

\begin{document}


\newcommand{\ofour}{PSR~J0437$-$4715}
\newcommand{\oo}{PSR~J0030$+$0451}
\newcommand{\fourteen}{J141256.0$+$792204}

\title{Measuring Neutron Star Radii via Pulse Profile Modeling with
  NICER}

\author{Feryal \"Ozel\altaffilmark{1}, Dimitrios
  Psaltis\altaffilmark{1}, Zaven Arzoumanian\altaffilmark{2}, Sharon
  Morsink\altaffilmark{3}, and Michi Baub\"ock\altaffilmark{1}}

\altaffiltext{1}{Astronomy Department, University of Arizona, 933 N.
  Cherry Ave., Tucson, AZ 85721, USA; fozel@email.arizona.edu}
\altaffiltext{2}{Center for Research and Exploration in Space Science and 
Technology \& X-Ray Astrophysics Laboratory, NASA Goddard Space Flight Center, 
Code 662, Greenbelt, MD 20771, USA}
\altaffiltext{3}{Department of Physics, University of
Alberta, 11455 Saskatchewan Drive, Edmonton, AB T6G 2E1, Canada}

\begin{abstract}
  The Neutron-star Interior Composition Explorer (NICER) is an X-ray
  astrophysics payload that will be placed on the International Space
  Station. Its primary science goal is to measure with high accuracy
  the pulse profiles that arise from the non-uniform thermal surface
  emission of rotation-powered pulsars. Modeling general relativistic
  effects on the profiles will lead to measuring the radii of these
  neutron stars and to constraining their equation of state. Achieving
  this goal will depend, among other things, on accurate knowledge of
  the source, sky, and instrument backgrounds. We use here simple
  analytic estimates to quantify the level at which these backgrounds
  need to be known in order for the upcoming measurements to provide
  significant constraints on the properties of neutron stars. We show
  that, even in the minimal-information scenario, knowledge of the
  background at a few percent level for a background-to-source
  countrate ratio of 0.2 allows for a measurement of the neutron star
  compactness to better than 10\% uncertainty for most of the
  parameter space.  These constraints improve further when more
  realistic assumptions are made about the neutron star emission and
  spin, and when additional information about the source itself, such
  as its mass or distance, are incorporated.
\end{abstract}

\keywords{stars: neutron --- dense matter -- equation of state -- pulsars: general --- gravitation}

\section{Introduction}

Periodic brightness oscillations originating from the surface of a
neutron star provide a powerful means of measuring its mass and radius
(see Strohmayer 2004 for an early review). In the case of
rotation-powered pulsars, the temperature non-uniformity that causes
the brightness oscillations is due to the magnetic field topology of
the stellar surface and the associated energetic particle flux from
the magnetosphere, which heats the polar caps relative to the rest of
the star such that they emit in the soft X-ray band (e.g., Ruderman \&
Sutherland 1975; Arons 1981; Harding \& Muslimov 2001, 2002; Bai \&
Spitkovsky 2010a, b). The trajectories of the photons emitted from the
surface are bent by the strong gravitational field of the neutron
star, imparting on the pulse shapes and amplitudes information about
the stellar compactness (e.g., Pechenick et al.\ 1983; Weinberg et
al.\ 2001; Poutanen \& Beloborodov 2006; Cadeau et al.\ 2007; Psaltis
\& \"Ozel 2014).

The degree to which this information can be decoded to infer the mass
and radius, either jointly or separately, depends on a number of
factors such as the X-ray brightness of the neutron star, its spin
period, $P$, the beaming pattern of the radiation emitted from the
surface, as well as the spot colatitude $\theta_s$ and the inclination
of the observer $i$ with respect to the stellar spin axis. This is
because all of these variables, along with the stellar compactness,
determine the pulse amplitude and its shape (i.e., the amplitudes of
the harmonics). In addition, there are several sources of background
that affect the measurement of the properties of the pulse profiles,
such as X-ray emission from the magnetosphere or the rest of the
neutron star surface, X-ray emission from other sources in the field
of view, and instrumental background. The resulting
signal-to-background ratio ultimately determines the accuracy with
which the stellar compactness, $u=2GM/Rc^2$, can be measured for a
star of mass $M$ and radius $R$.

NICER is a NASA mission, currently in development, designed primarily
to observe thermal X-rays originating from the polar caps of a number
of rotation-powered millisecond pulsars. Its goal is to analyze pulse
profiles to measure the stellar compactness and radius (Gendreau et
al.\ 2012; Arzoumanian et al.\ 2014). Being a dedicated mission, it
will have the ability to observe selected sources for long periods to
accumulate the necessary number of counts to measure the properties of
the pulse profile. Nevertheless, the accuracy it will achieve in the
measurement of the various harmonic amplitudes, and, thus, in the
compactness, will depend on the level and our knowledge of the several
sources of background. Because it is not an imaging instrument and
because of the International Space Station environment, it will
register both time-varying and constant sources of background, arising
from the various origins discussed above. In this paper, we formulate
the effect of the background on the pulse profiles and investigate the
background tolerances that will still enable 5-10\% accuracy in the
measurements of the neutron star properties, which is necessary to pin
down the dense matter equation of state. We provide approximate
formulae for the uncertainties in the inferred stellar compactness and
radius as a function of the signal-to-noise ratio, the uncertainty in
the measurement of the background, and the uncertainty in the system
geometry, described by the spot colatitude and the observer's
inclination.

Thermal emission from the polar caps of rotation powered pulsars has
been observed with previous generations of X-ray satellites (\"Ogelman
1991; Finley et al.\ 1992; Becker \& Tr\"umper 1993; Pavlov \& Zavlin
1997). Pulsed X-ray emission has been detected and the thermal X-ray
brightness has been determined for a number of sources, such as
\ofour, \oo, PSR~J2124$-$3358, and PSR~J1024$-$0719, most recently by
targeted Chandra and XMM-{\it Newton} observations (e.g., Zavlin 2006;
Bogdanov et al.\ 2007; Bogdanov \& Grindlay 2009; Bogdanov 2013). In
addition, the number of rotation-powered millisecond pulsars has grown
exponentially in recent years, thanks to discoveries by the {\it
  Fermi} satellite (e.g., Abdo et al.\ 2013) and radio
surveys\footnote{See
  \url{https://confluence.slac.stanford.edu/display/GLAMCOG/Public+List+of+LAT-Detected+Gamma-Ray+Pulsars}
  and \url{http://astro.phys.wvu.edu/GalacticMSPs} for up-to-date
  lists.}, some of which are likely to exhibit pulsed soft X-ray
emission. The spectra are usually described by a thermal component
originating from the surface, typically modeled by unmagnetized
hydrogen atmospheres or blackbodies, as well as a non-thermal
component, likely of magnetospheric origin, which manifests itself as
a power-law tail at higher energies. Using this pulsed surface
emission to constrain the neutron star properties has so far led to
broad constraints on masses and radii (Zavlin 2006; Bogdanov et
al.\ 2007; Bogdanov \& Grindlay 2009; Bogdanov 2013).

The rotation-powered pulsars that are the primary targets for NICER
have spin frequencies of $200-400$~Hz. This range of relatively low
frequencies (compared to neutron stars in some accreting systems)
minimizes higher order spin effects such as the stellar oblateness and
the spacetime quadrupole (Morsink et al.\ 2007; Cadeau et al.\ 2007;
Psaltis \& \"Ozel 2014; Psaltis et al.\ 2014). The first order Doppler
effects can still be appreciable in this range of spin
frequencies. This simplifies the modeling by allowing us to treat the
stellar spacetime in the Schwarzschild+Doppler approximation. A large
amount of work has been carried out in this regime, starting with the
numerical calculations of self-lensing in the Schwarzschild metric by
Pechenick et al.\ (1983) and its application to the millisecond
brightness oscillations observed by the Rossi X-ray Timing Explorer
(Strohmayer et al.\ 1997; Nath et al.\ 2002). Furthermore, Poutanen \&
Beloborodov (2006) devised an analytic expression that we use in the
present study when calculating the amplitudes of the harmonics.

In this paper, we investigate the effect of background on the
determination of the neutron star compactness given measurements of
the shapes of pulse profiles from rotation-powered
pulsars. Specifically, we make use of the analytic formulae for the
pulse shapes of moderately spinning neutron stars in the Schwarzschild
metric to calculate the number of observables one can determine from
the profiles in different situations. We then use these to quantify
the level at which the various sources of background must be
independently known in order for the upcoming NICER measurements to
provide significant constraints on the properties of neutron stars.

\section{Measuring the Properties of Pulse Profiles}

In the following, we will assume that the surface emission of a slowly
spinning, rotation powered pulsar originates in one small circular
polar cap on its surface. Baub\"ock et al.\ (2015) estimated the
typical sizes of polar caps, defined as the radius within which the
open magnetic field lines originate from the neutron star surface, for
the NICER targets.  The analytic estimates in that study of the size
of the polar cap (Baub\"ock et al.\ 2015), as well as numerical
simulations of surface heating caused by return magnetospheric
currents (e.g., Bai \& Spitkovsky 2010), support the assumption that
the polar cap size is too small for its detailed emission structure to
affect the pulse profile observed at infinity.  Because of this, we
make the small spot approximation when calculating the harmonic
amplitudes of the pulse profiles.  However, observations of the
prototypical NICER target, PSR~J0437$-$4715, suggest that its surface
emission may arise from two small but non-antipodal polar caps
(Bogdanov 2014). This latter case is more complicated than what we
explore here and will change the actual shape of the pulse profile,
but its complexity does not severely affect our conclusions in the
uncertainties achieved in the various quantities as a function of
count rate and backgrounds.

For a slowly spinning neutron star, the pulse profile observed by
NICER will be characterized by the following quantities (see
Figure~1): the background countrate $A$, the constant component of the
pulse profile $Q$, the amplitudes $V_n$ and the phases $\phi_n$ of the
various harmonics, and, if present, the phase $\phi_{\rm p}$ beyond
which the flux from the polar cap is blocked by the star. The general
form of the profile will be
\begin{equation}
F(t)=A+\left\{\begin{array}{ll}
Q+\sum_{n=1} V_n\cos(\frac{2\pi n t}{P} + \phi_n), &\mbox{if}~\vert 2\pi t/P\vert<\phi_{\rm p}\\
0, &\mbox{otherwise}
\end{array}\right.
\label{eq:profile}
\end{equation}
If the spin of the neutron star is significant, then additional 
harmonics appear in the profile with orthogonal phases.

\begin{figure*}[t]
  \psfig{file=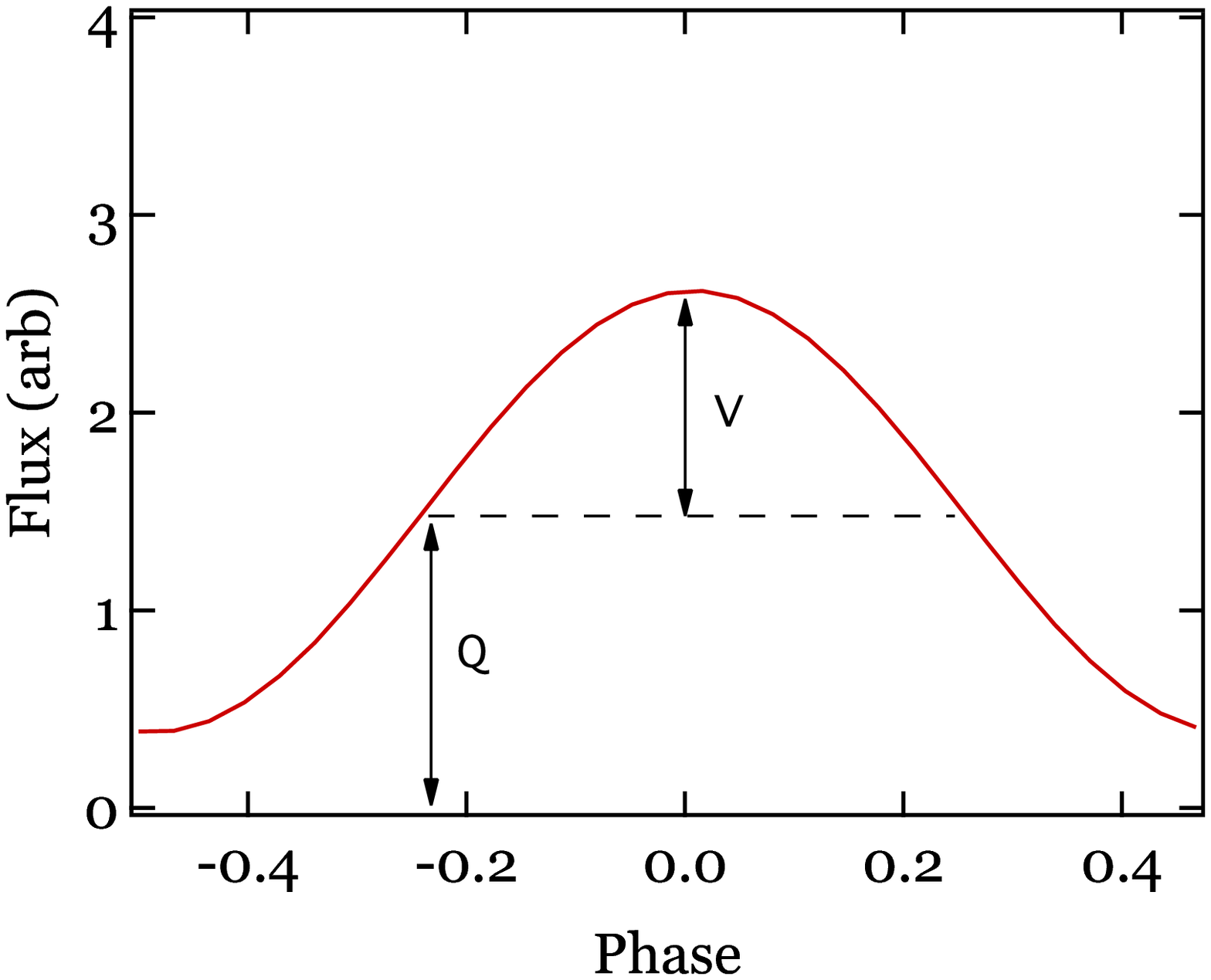,width=3.5in,clip=}
  \psfig{file=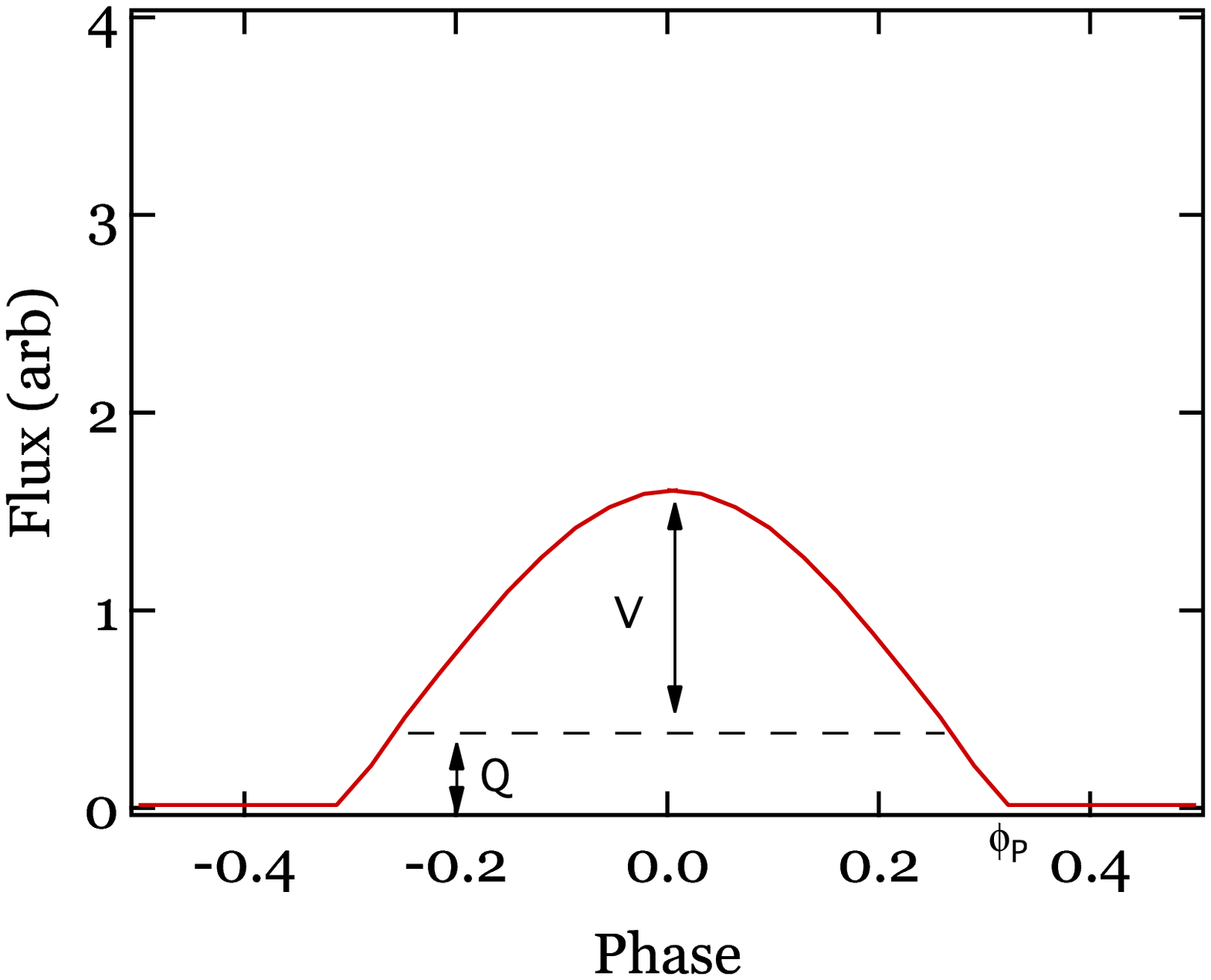,width=3.5in,clip=}
  \psfig{file=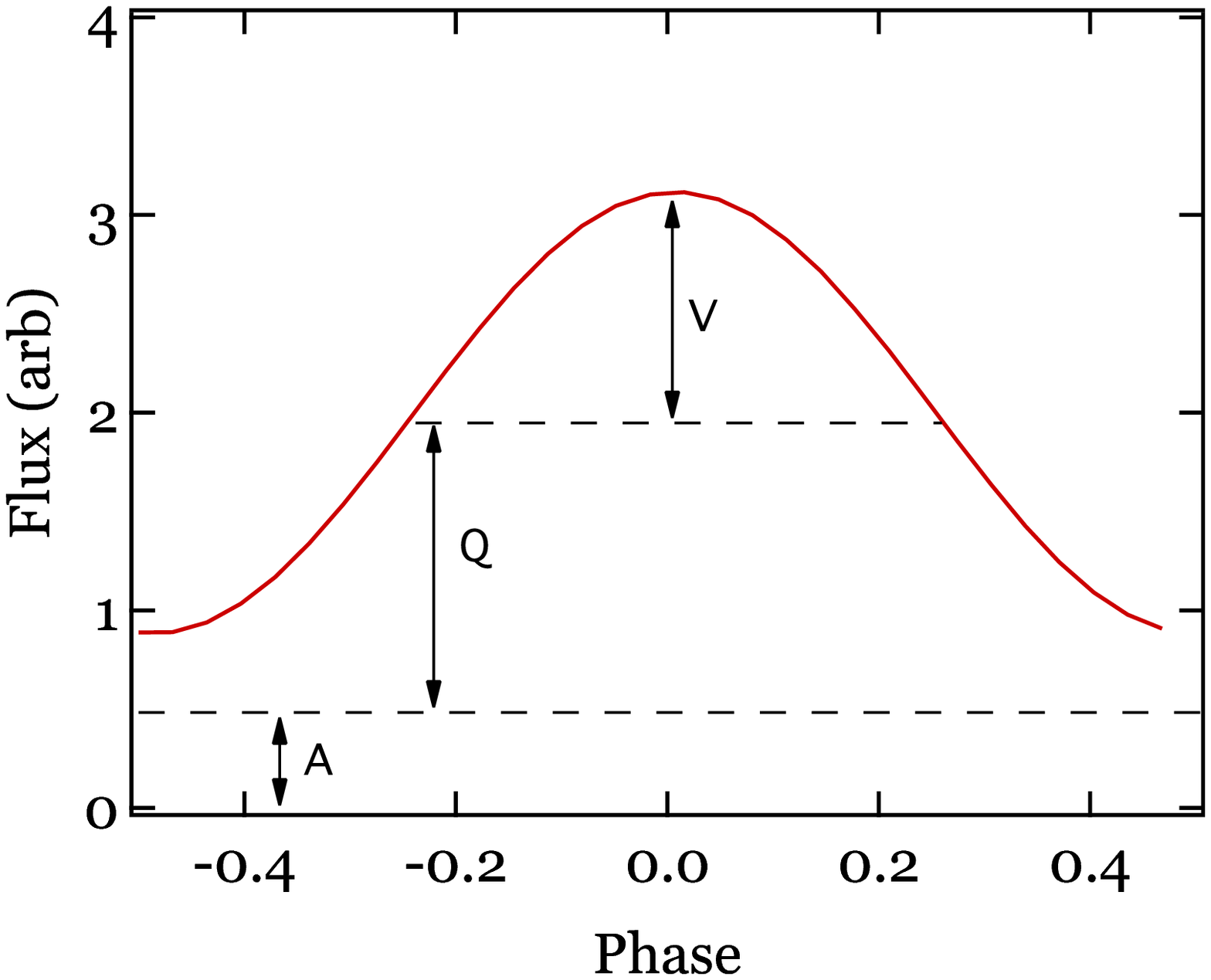,width=3.5in,clip=}
  \psfig{file=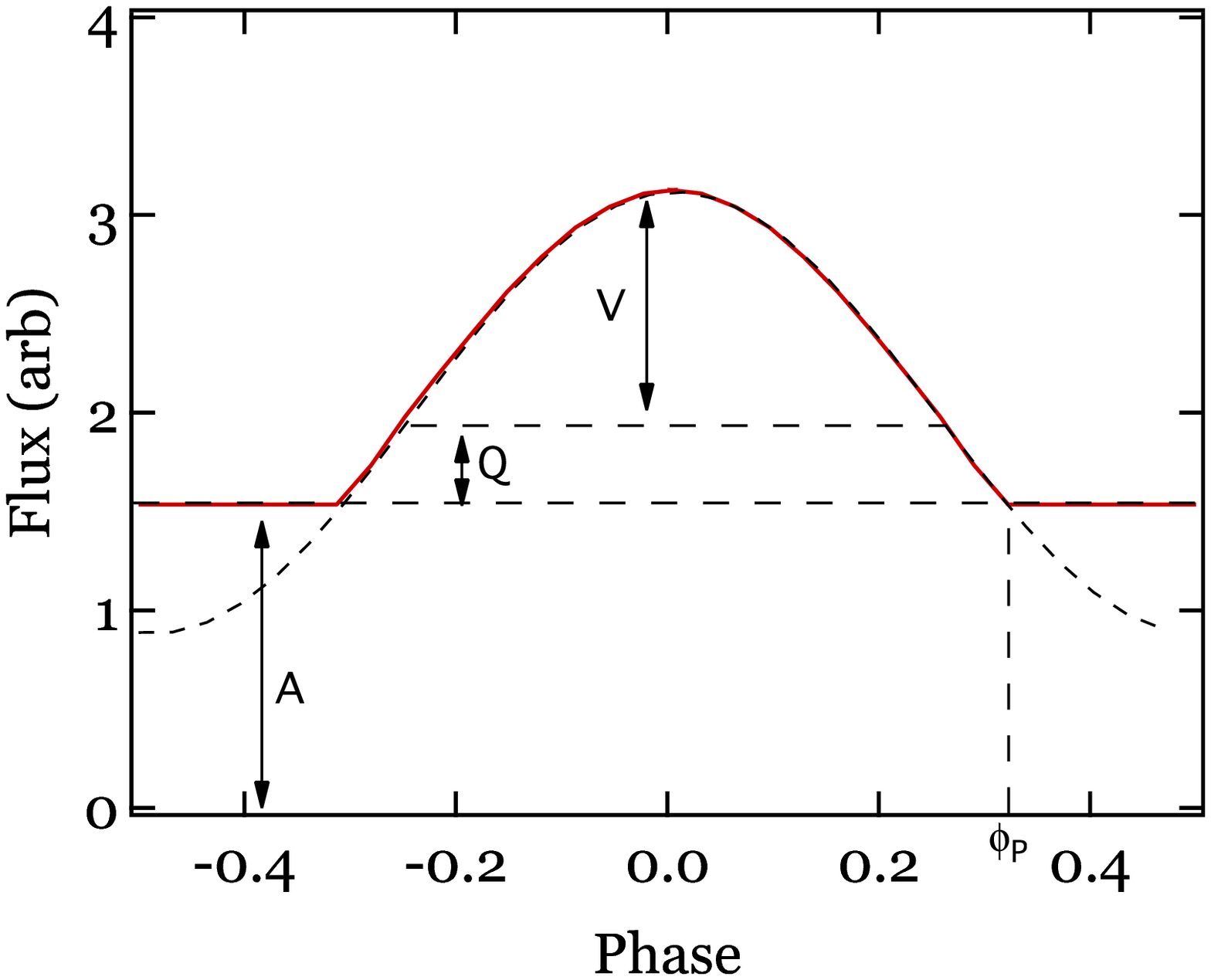,width=3.5in,clip=}
\caption{Pulse profiles generated by a circular hot spot on the
  surface of a slowly spinning neutron star, without {\em (top\/}) and
  with {\em (bottom}) instrument or sky background. The left panels
  show a situation in which the polar cap is visible throughout the
  pulse period, whereas the right panels show the pulse shape when the
  polar cap is hidden behind the star for part of the period. In
  general, a profile is characterized by four quantities: the constant
  component of the pulse profile $Q$ (see Eq.~\ref{eq:profile}), the
  amplitude of the sinusoidal profile $V$ (assumed for the purposes of
  this figure to have only a non-zero fundamental component), the
  background countrate $A$, and, potentially, the phase $\phi_{\rm p}$
  beyond which the polar cap is not visible to the observer. Note that
  pulse phase in the interval $[-0.5-0.5]$ is shown here, but radians
  are used in the equations and text.}
\label{fig:profile_occult}
\mbox{}
\end{figure*}

The total number of counts collected when $N$ periods are accumulated 
is equal to
\begin{eqnarray}
  C&=&N\int_0^PF(t)dt\nonumber\\
  &=&(A+Q\frac{\phi_{\rm p}}{\pi})NP 
  +2N\sum_{n=1}V_n\int_0^{P\phi_{\rm p}/2\pi}\cos\left(\frac{2\pi n t}{P}+\phi_n
  \right)dt\nonumber\\
  &=&(A+Q\frac{\phi_{\rm p}}{\pi})NP
  +\sum_{n=1}\frac{V_nNP}{\pi}
\left[\sin(n \phi_{\rm p} + \phi_n) - \sin \phi_n \right]\;.
\end{eqnarray}
The uncertainty in the measurement of the total number of counts will
simply be 
\begin{equation}
\sigma_{\rm C}=\sqrt{C}\simeq \sqrt{(A+Q)NP}\;,
\end{equation}
where, in the last equality, we assumed that the polar cap is visible
throughout the pulse period (i.e., $\phi_{\rm p} = \pi$). From this
last equation, we also get the uncertainty in measuring the sum $Q+A$
as
\begin{equation}
\sigma_{Q+A}=\frac{\sigma_c}{NP}=\sqrt{\frac{A+Q}{NP}}\;.
\end{equation}

In the following calculation, we will need primarily the constant
component of the source countrate $Q$ and the associated uncertainty
$\sigma_{\rm Q}$. In order to do this, we will need a measurement of
the background countrate, $A$, and its associated uncertainty,
$\sigma_A$. The latter will have a Poisson component that will depend
on the total time $T$ used to measure the background, as well as a
possible systematic component, such as
\begin{equation}
\sigma_{\rm A}=\sqrt{\frac{A}{T}}+\lambda A
\end{equation}
with $\lambda$ an appropriate constant. With these definitions, we
obtain
\begin{eqnarray}
\frac{\sigma_{\rm Q}}{Q}&=&\frac{\left(\sigma_{Q+A}^2+\sigma_A^2\right)^{1/2}}
{Q}\nonumber\\
&=&\left[\left(\frac{A}{Q}+1\right)\frac{1}{QNP}+
\left(\frac{\sigma_A}{A}\right)^2\left(\frac{A}{Q}\right)^2\right]^{1/2}\;.
\end{eqnarray}
We can also write this equation in terms of the total source counts
$S=QNP$ and the total background counts $B=ANP$ as
\begin{equation}
\frac{\sigma_{\rm Q}}{Q}=\left[\left(\frac{B}{S}+1\right)\frac{1}{S}+
  \left(\frac{\sigma_B}{B}\right)^2\left(\frac{B}{S}\right)^2\right]^{1/2}\;.
\label{eq:sigmaQ}
\end{equation}

\begin{figure}[t]
  \psfig{file=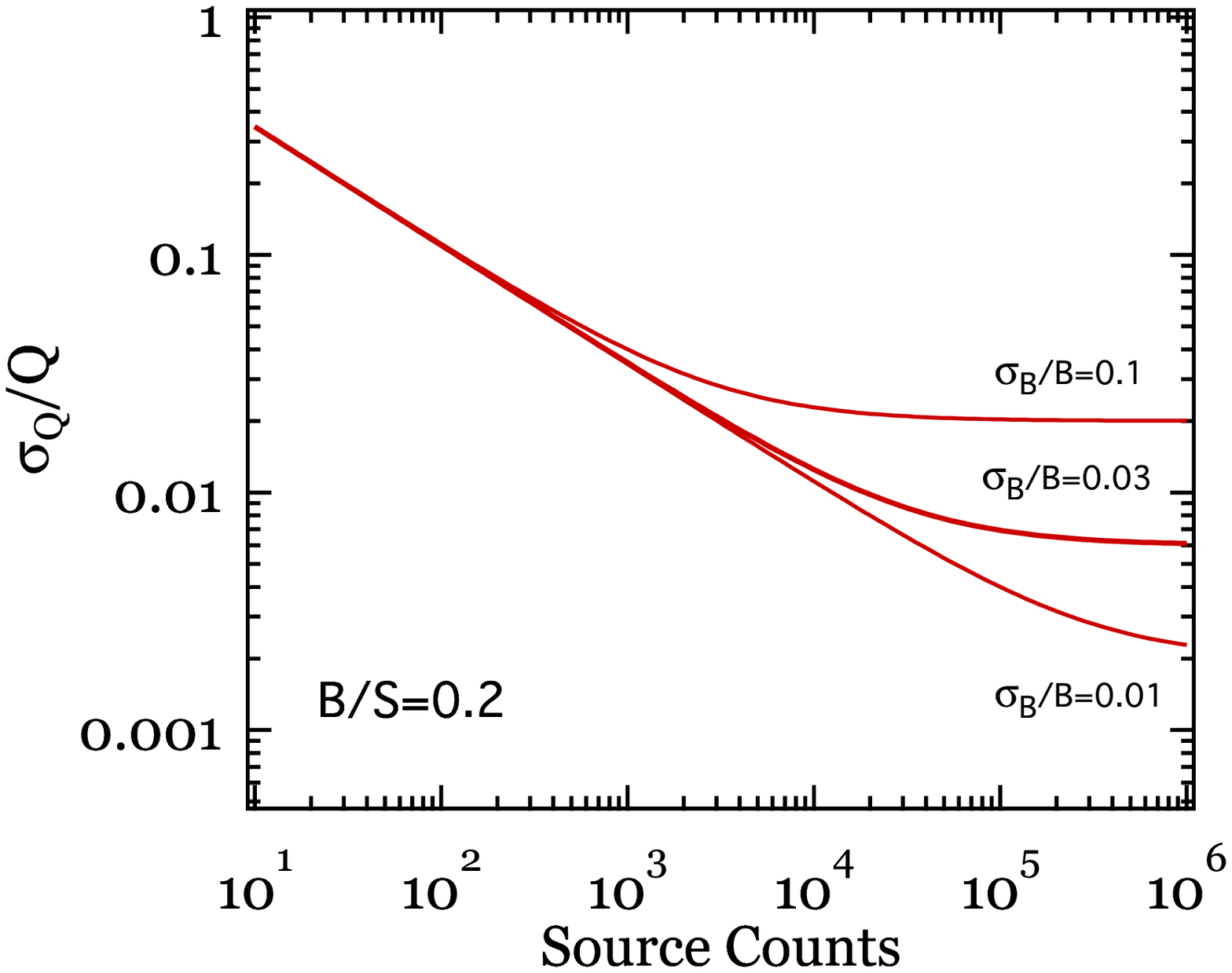,width=3.5in,clip=}
  \psfig{file=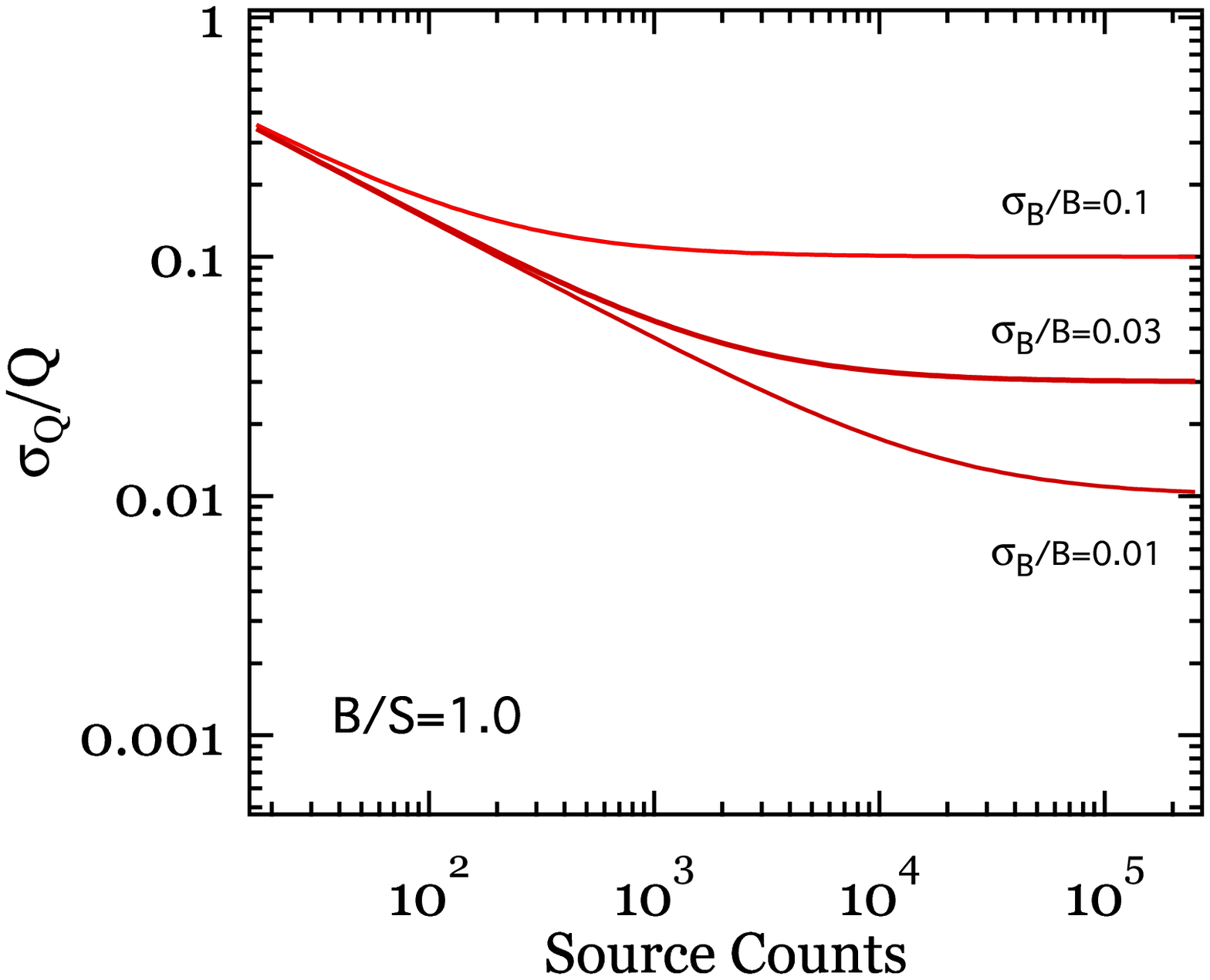,width=3.5in,clip=}
\caption{Predicted fractional uncertainties in the measurement of the
  quantity $Q$, i.e., the constant component, for a nearly sinusoidal pulse
  profile, as a function of the total number of source counts
  collected. The two panels correspond to two different ratios of the
  background to source counts and different curves in each panel are
  labeled by the fractional uncertainty in the knowledge of the
  background counts $\sigma_B/B$.}
\label{fig:sigmaq}
\mbox{}
\end{figure}

Figure~\ref{fig:sigmaq} shows the predicted fractional uncertainties in 
the measurement of the quantity $Q$ as a function of the total number
of source counts collected. In making this figure, we have assumed a 
constant fractional uncertainty in the {\em a priori\/} knowledge
of the background, i.e., $\sigma_B/B$, and a constant ratio of
background to source counts, i.e., $B/S$. Under these assumptions,
the asymptotic value of the uncertainty (i.e., for a large number of 
source counts) is dominated by the uncertainty in the knowledge
of the background and scales as
\begin{equation}
\lim_{S\rightarrow \infty}\frac{\sigma_{\rm Q}}{Q}=\left(\frac{\sigma_B}{B}\right)
\left(\frac{B}{S}\right)\;.
\end{equation}
In other words, the fractional uncertainty in the measurement of $Q$
is limited by the product of the fractional uncertainty in the knowledge
of the background times the ratio of the background to source counts.

The additional quantities that we can measure from the pulse profile
are the fractional r.m.s.\ amplitudes of the various
harmonics. Considering, for example, the $n-$th harmonic, the measured
fractional r.m.s.\ amplitude (including the background) is
\begin{equation}
r_{\rm obs}=\frac{V_n}{\sqrt{2}(Q+A)}\;.
\end{equation}
Using the result in Psaltis et al.\ (2014), the uncertainty
in the measurement of $r_{\rm obs}$ scales as
\begin{equation}
\sigma_{\rm r,obs}=\frac{\sqrt{S+B}}{S}=\frac{\sqrt{(A+Q)NP}}{QNP}\;,
\end{equation}
from which we can write
\begin{eqnarray}
\sigma_V^2&=&\left(\frac{\partial V_n}{\partial r_{\rm obs}}\right)^2 \sigma_{\rm r,obs}^2
+\left[\frac{\partial V_n}{\partial (Q+A)}\right]^2 \sigma_{Q+A}^2\nonumber\\
&=&\frac{2(Q+A)^3}{Q^2NP}+\frac{V_n^2}{(Q+A)NP}\;.
\end{eqnarray}

\begin{figure}[t]
  \psfig{file=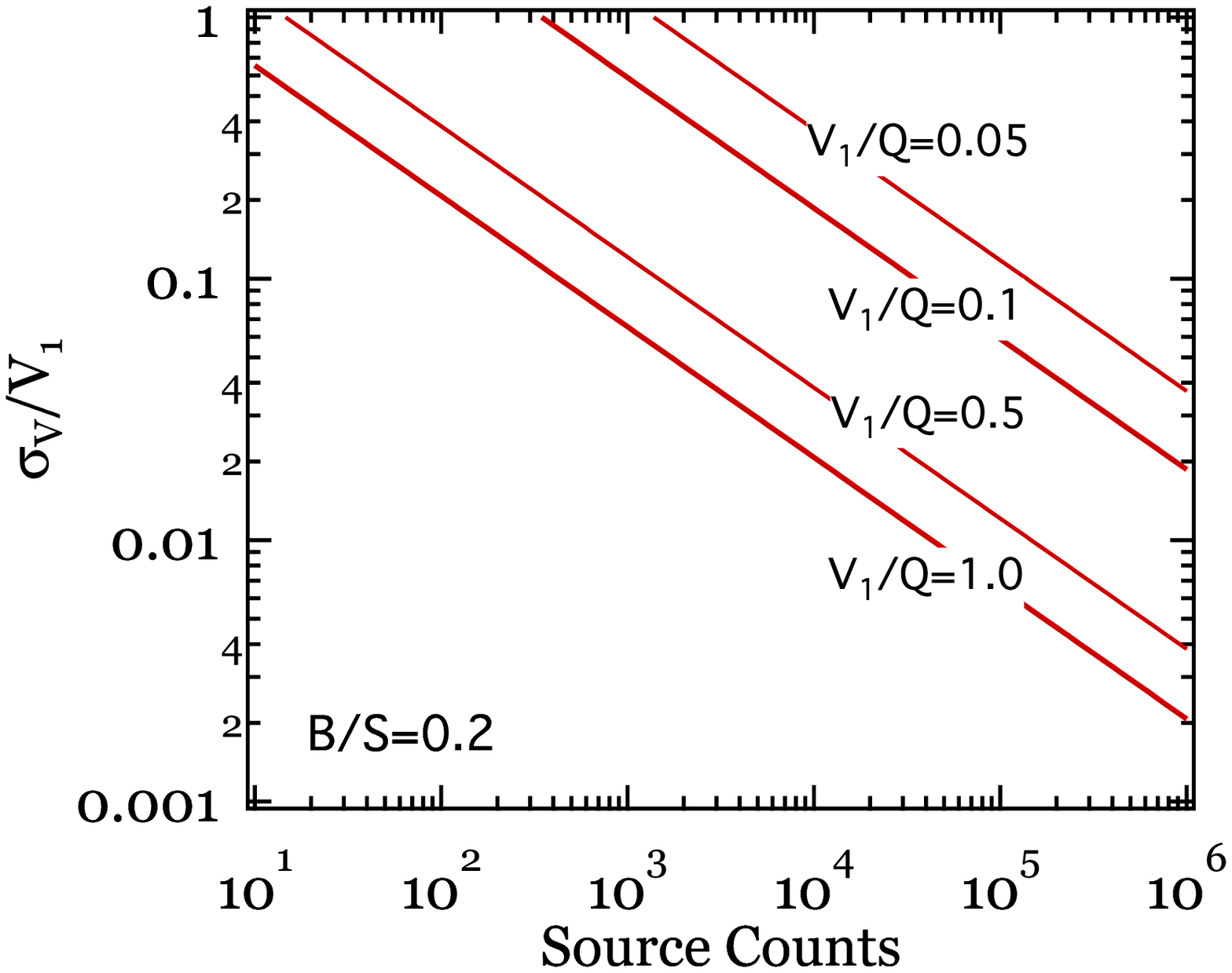,width=3.5in,clip=}
  \psfig{file=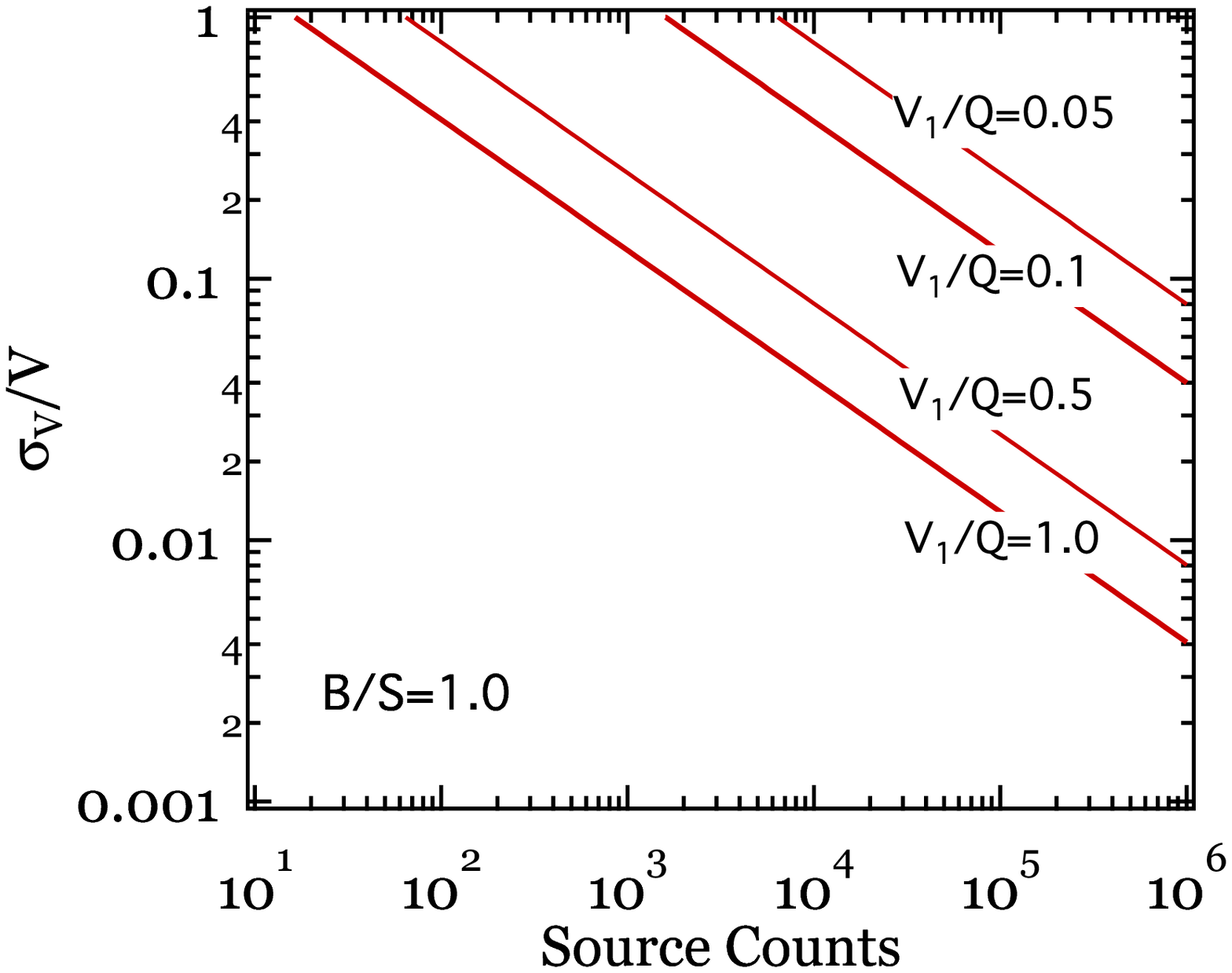,width=3.5in,clip=}
\caption{Predicted fractional uncertainties in the measurement of the
  quantity $V_1$, i.e., the amplitude of the fundamental, for a nearly
  sinusoidal pulse profile, as a function of the total number of
  source counts collected. The two panels correspond to different
  ratios of the background to source counts and different curves in
  each panel are labeled by the intrinsic fractional amplitude of the
  pulsations $V_1/Q$.}
\label{fig:sigmav}
\mbox{}
\end{figure}

The fractional uncertainty in the measurement of the amplitude is
\begin{equation}
\frac{\sigma_V}{V_n}=\left[2\left(1+\frac{A}{Q}\right)^3
  \left(\frac{Q}{V_n}\right)^2\frac{1}{QNP}+\left(1+\frac{A}{Q}\right)^{-1}
\frac{1}{QNP}\right]^{1/2}\;.
\end{equation}
Writing this equation in terms of the total number of source and
background counts, we obtain
\begin{equation}
\frac{\sigma_V}{V_n}=\left[2\left(1+\frac{B}{S}\right)^3
  \left(\frac{Q}{V_n}\right)^2+\left(1+\frac{B}{S}\right)^{-1}
  \right]^{1/2}\frac{1}{S^{1/2}}\;.
\label{eq:sigmaV}
\end{equation}

Figure~\ref{fig:sigmav} shows the predicted fractional uncertainty in
the measurement of the quantity $V$, as a function of the total number
of source counts. As the last equation shows, the uncertainty in the
measured amplitude of pulsations simply scales as the inverse square
root of the source counts.

Finally, the last quantity to be measured, when present in the pulse
profile, is the phase $\phi_{\rm P}$, beyond which the polar cap is
blocked by the star. The main uncertainty in this measurement will
arise from the fact that the evolution of the pulse profile with phase
is masked by the uncertainty in the background, i.e.,
\begin{equation}
V_1\cos\left(\phi_{\rm P}+\delta \phi_{\rm P}\right)\simeq \sigma_{\rm A}\;.
\end{equation}
Setting $\sigma_{\phi}\equiv \delta \phi_{\rm P}$, we can use this
equation to find
\begin{equation}
\sigma_\phi=\frac{\sigma_A}{V_1\sin\phi_{\rm p}}\;.
\end{equation}
and write this last equation in the more useful form
\begin{equation}
\sigma_\phi=\frac{1}{\sin\phi_{\rm p}}
\left(\frac{\sigma_B}{B}\right)
\left(\frac{B}{S}\right)
\left(\frac{Q}{V_1}\right)\;.
\end{equation}
For the remainder of our discussion, we will only consider pulse
shapes in which the polar cap is visible throughout the pulse period.

\section{From pulse profiles to stellar compactness}

Having measured the characteristic properties of a pulse profile,
i.e., its constant component $Q$ and the amplitudes $V_n$ and phases
$\phi_n$ of its various harmonics, we can use theoretical models of
the gravitationally lensed emission from a neutron-star hotspot to
obtain the compactness and the radius of the neutron star.

In the following, we will make use of the approximate analytic
expressions and terminology of Poutanen \& Beloborodov (2006).  These
expressions take into account light bending around a slowly spinning
star and yield the bolometric flux observed at infinity under the
assumption that the polar cap is infinitesimally small and that no
emission emerges from the rest of the star. We will denote by $M$ and
$R$ the mass and radius of the neutron star, by $P$ its spin period,
by $i$ the inclination of the observer with respect to the spin axis,
and by $\theta$ the colatitude of the hot spot. 

Model atmospheres of neutron stars, in general, and of magnetic polar
caps, in particular, show that the emission from the neutron-star
surface is not isotropic but is beamed (see, e.g., the discussion in
\"Ozel 2013). We will represent this beaming with a functional form
\begin{equation}
I(\alpha)=I_0(1+h\cos\alpha)\;,
\end{equation}
where $\alpha$ is the emission angle with respect to the normal to the
surface and $h$ is a quantity that is calculated from model
atmospheres and measures the degree of beaming ($h=0$ being the
isotropic limit).

We will also define the compactness of the neutron star as
\begin{equation}
  u\equiv\frac{2GM}{Rc^2}
\end{equation}
as well as the following two quantities
\begin{eqnarray}
q&\equiv&u+(1-u)\cos i\cos\theta\nonumber\\
v&\equiv&(1-u)\sin i\sin\theta\;.
\label{eq:ns_prop}
\end{eqnarray}
The relations between the characteristics of the pulse profile (see
eq.~[\ref{eq:profile}]) and the properties of a slowly spinning
neutron star are (Poutanen \& Beloborodov 2006, eq.~[28]-[38])
\begin{eqnarray}
Q&=&I_0{\cal G}\frac{dS}{D^2}
\left[q+h\left(q^2+\frac{v^2}{2}\right)\right]\nonumber\\
V_1&=&I_0{\cal G}\frac{dS}{D^2}
\frac{(1+2hq)v}{\cos \phi_1}  \nonumber\\
V_2&=&I_0{\cal G}\frac{dS}{D^2} \frac{hv^2}{2 \cos \phi_2} \nonumber \\
\tan \phi_1 &=& \frac{(3+\Gamma)q+(4+\Gamma)h(q^2+v^2/4)}{u(1+2hq)} 
\left(\frac{2\pi R}{cP}\right) \sin i \sin \theta \nonumber\\
\tan \phi_2 &=& \frac{(4+\Gamma)(1+2hq)-1}{hu} 
\left(\frac{2\pi R}{cP}\right) \sin i \sin \theta \;.
\label{eq:general}
\end{eqnarray}
In these equations, $dS$ is the surface area of the polar cap, $D$ is
the distance to the source, $\Gamma$ is the photon index of the
spectrum, and ${\cal G}$ is a (calculable) factor that depends on the
spectrum of the source and the photon-energy range of the detector.

\subsection{The Limiting Case of Minimal Information}

To calculate the background tolerances in the measurements of the
neutron star properties from the characteristics of its pulse profile,
we will first consider the worst possible case, i.e., one in which the
emission from the neutron star surface is isotropic and the Doppler
effects on the pulse profile are not measurable. Although this is not
expected in reality, this situation both reduces the amplitude of the
fundamental (exacerbating background effects) and the amplitudes of
the harmonics, thus minimizing the number of measurable quantities in
the profile. This can be seen by taking the limit $h=0$ and $2\pi R/cP
\rightarrow 0$ in the above equations, which simplifies them to
\begin{eqnarray}
Q&=&I_0{\cal G}\frac{dS}{D^2}q
\nonumber\\
V_1&=&I_0{\cal G}\frac{dS}{D^2}v\nonumber\\
V_2&=&0\;.
\end{eqnarray}
One of the most significant benefits of neutron-star radius
measurements with pulse profile modeling is that, in principle, they
can provide results that are independent of any prior knowledge of the
distance to each source (although, as we will discuss below, when
precise distance measurements are available, they can be folded in to
increase the number of independently measured quantities from the
pulse profile). In order not to rely on the distance information, we
cannot separately use the absolute constant component and fundamental
amplitude of the pulse profile (both of which depend on the size of
the polar cap and on the distance), but we can only consider the
fractional amplitude of the fundamental, i.e.,
\begin{equation}
  r_1\equiv\frac{V_1}{Q}=\frac{v}{q}=\frac{(1-u)\sin i\sin\theta}{
    u+(1-u)\cos i\cos\theta}\;.
  \label{eq:r1}
\end{equation}
In this last step, we used the definitions~(\ref{eq:ns_prop}) to
connect the pulse profile properties to the neutron-star compactness.
Solving this last equation for the compactness $u$, we obtain
\begin{equation}
  u=\frac{r_1\cos i\cos\theta-\sin i\sin\theta}
  {-r_1+r_1\cos i\cos\theta -\sin i\sin\theta}\;.
\end{equation}

Clearly, in order to measure the neutron-star compactness in this
case, we need independent knowledge of the inclination of the
observer, $i$, and of the colatitude of the polar cap, $\theta$.  Such
prior knowledge is not unreasonable for many rotation-powered pulsars,
the geometries of which can be inferred by fitting model $\gamma$-ray
pulse profiles to {\it Fermi} observations of individual sources
(Johnson et al.\ 2014) and, to a certain extent, by modeling
polarization angle swings in the radio (Yan et al.\ 2011; Dai et
al.\ 2015).

The uncertainty within which the fractional amplitude of the fundamental
can be measured is
\begin{equation}
  \frac{\sigma_{r1}}{r_1}=\left(
  \frac{\sigma_{V1}^2}{V_1^2}+
      \frac{\sigma_{Q}^2}{Q^2}\right)^{1/2}\;,
\end{equation}
where $\sigma_Q/Q$ and $\sigma_{V1}/V_1$ are given by
equations~(\ref{eq:sigmaQ}) and (\ref{eq:sigmaV}), respectively. At the
limit of a very large number of counts,
\begin{equation}
\lim_{S\rightarrow \infty}\frac{\sigma_{r1}}{r_1}=\left(\frac{\sigma_B}{B}\right)
\left(\frac{B}{S}\right)\;.  
\end{equation}
Using these expressions, we can finally write the uncertainty in the
measurement of the neutron-star compactness as
\begin{eqnarray}
  \sigma_u&=&\left[
    \left(\frac{\partial u}{\partial r_1}\right)^2\sigma_{r1}^2
    +\left(\frac{\partial u}{\partial i}\right)^2\sigma_{i}^2
    +\left(\frac{\partial u}{\partial \theta}\right)^2\sigma_{\theta}^2
    \right]^{1/2}\nonumber\\
  &=&
  \left[\sigma_{r1}^2 \sin^2i\sin^2\theta\right.
  +\frac{1}{2}r_1^3 \sin (2 i)\sin(2\theta)
    \left(\sigma_i^2+\sigma_\theta^2\right)
    \nonumber\\
     &&
   +r_1^2 \cos^2i\sin^2\theta
   \left(\sigma_i^2+r_1^2 \sigma_\theta^2\right)\nonumber\\
   &&\left.
   +r_1^2 \sin ^2i\cos ^2\theta
   \left(r_1^2 \sigma_i^2+\sigma_\theta^2\right)\right]^{1/2}\nonumber\\
     &&
  \left(r_1-r_1 \cos i \cos \theta +\sin i \sin\theta\right)^{-2}\;.
\end{eqnarray}
Here, $\sigma_i$ and $\sigma_\theta$ are the uncertainties (in radians)
in the prior knowledge of the observer inclination and of the colatitude
of the polar cap.

\begin{figure}[t]
  \psfig{file=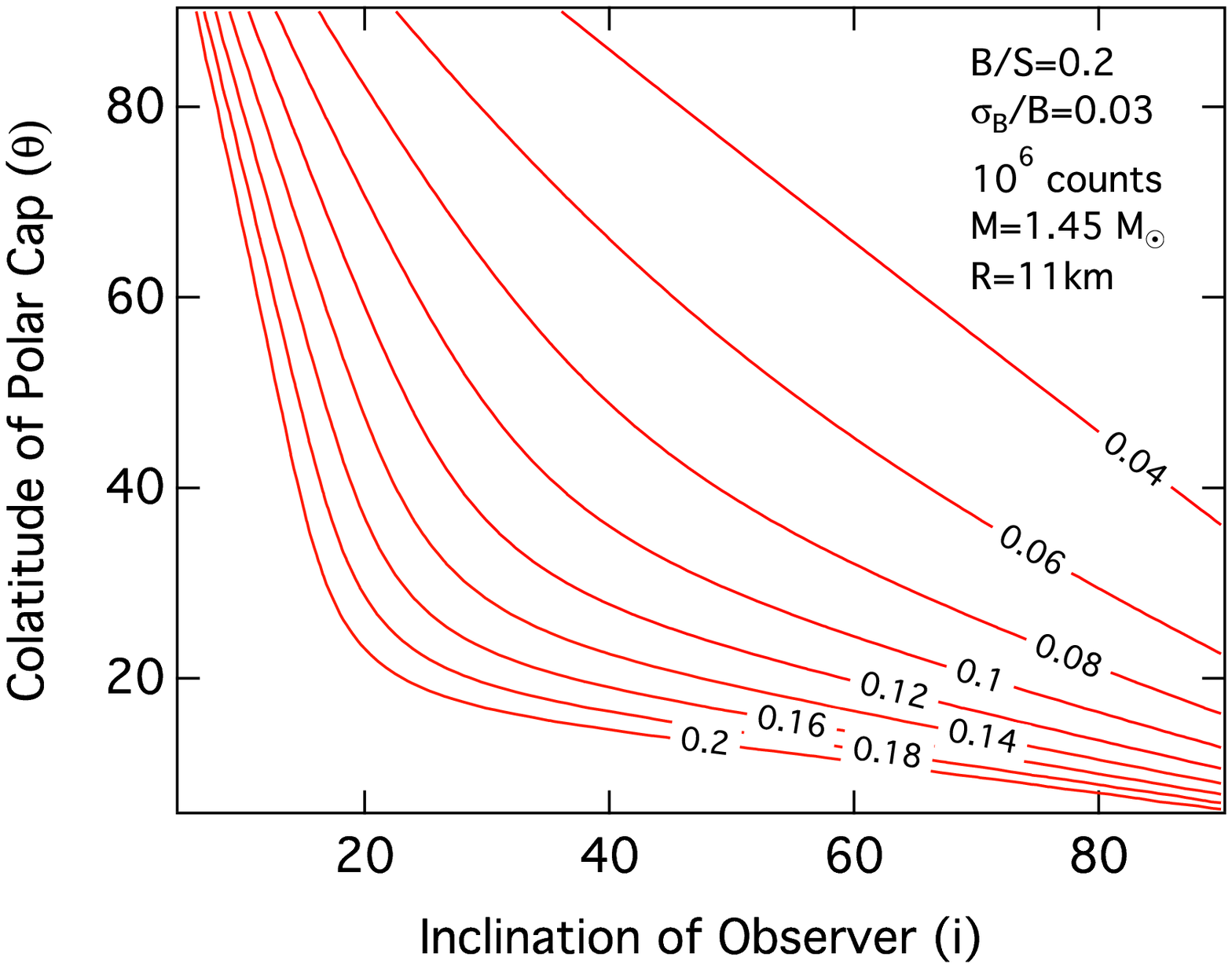,width=3.5in,clip=}
  \psfig{file=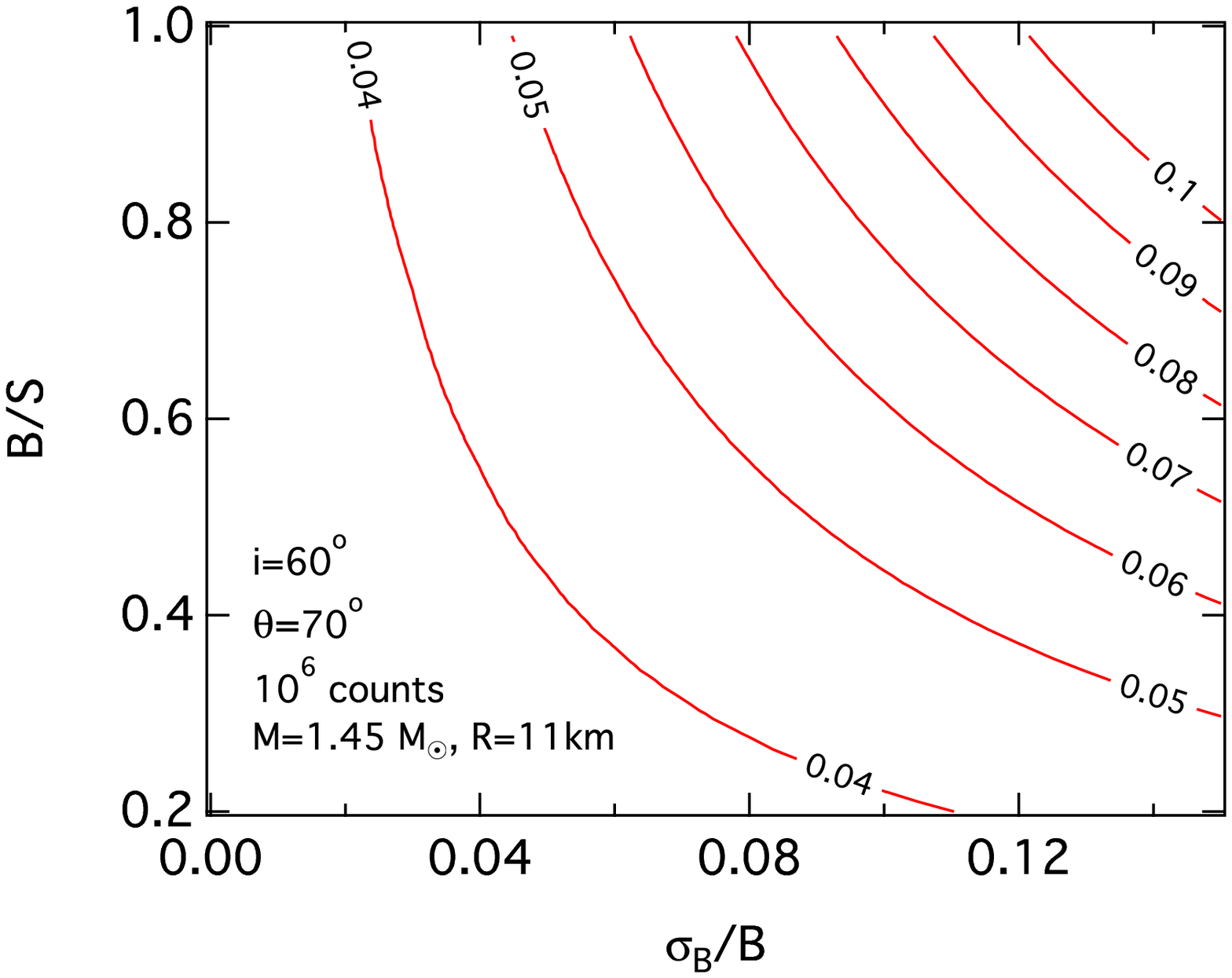,width=3.5in,clip=}
  \caption{(Left) Contours of constant predicted fractional error in
    the measurement of a neutron-star compactness $u$, for different
    orientations of the observer $i$ and colatitudes of the polar cap
    $\theta$. For this set of calculations, we assumed a neutron star
    mass of $1.45 \; M_\odot$ and a radius of 11~km, as well as a
    total of $10^6$ source counts, a background-to-source count ratio
    $B/S=0.2$, and an uncertainty in the background $\sigma_B/B=0.03$.
    We also assumed that the two angles $i$ and $\theta$ to be known
    with an uncertainty of $\sigma_i=\sigma_\theta=2^\circ$. (Right)
    Contours of constant predicted fractional error in the measurement
    of a neutron-star compactness $u$, for different
    background-to-source count ratios $B/S$ and amount of uncertainty
    in the background $\sigma_B/B$. For this set of calculations, the
    inclination of the observer was set to $i=60^\circ$ and the
    colatitude of the polar cap was set to $\theta=70^\circ$. The
    remaining parameters are the same as in the left panel.}
\label{fig:angles}
\mbox{}
\end{figure}

Figure~\ref{fig:angles} shows the predicted fractional uncertainties
in the measurement of the compactness of a neutron star, for different
parameters of the system as well as for different degrees of knowledge
of the background to the source. When the geometric angles $i$ and
$\theta$ are reasonably well known via other measurements, these
figures show that knowledge of the background at a few percent level
allows a measurement of the neutron star compactness to better than
$10\%$ uncertainty for most of the parameter space. This result is
obtained for a background-to-source ratio of 0.2, which is achievable
for the brightest NICER targets, and a total accumulation of $10^6$
counts. It is only when the pulse amplitudes plummet for the cases
where either the observer or the polar cap are nearly aligned with the
spin axis that the uncertainties become larger than $15\%$ (see the
left panel of Fig.~\ref{fig:angles}). When the system geometry is such
that the pulse amplitudes are reasonably large (as is the case with
prototypical NICER targets), the right panel of
Figure~\ref{fig:angles} shows the dependence of the fractional error
in the neutron star compactness for different background-to-signal
ratios and different levels of uncertainty in the determination of the
background. Even when the background is comparable to the signal and
is measured only to within $\approx 15\%$, the fractional error in the
compactness remains at or below $10\%$. Note that this is for the
minimal-information case of isotropic beaming, where only one
fractional amplitude can be obtained from the pulse profiles.

We also calculate the limiting case of negligible uncertainties in the
knowledge of these angles and of an infinite number of source counts.
The uncertainty in the measurement of the neutron-star compactness
then becomes
\begin{equation}
  \lim_{S\rightarrow \infty}\sigma_u=\left(\frac{\sigma_B}{B}\right)
  \left(\frac{B}{S}\right)
  \frac{\vert\sin i\sin\theta\vert r_1}
       {\left(-r_1 \cos i \cos \theta +\sin i \sin\theta+r_1\right)^2}
  \;.  
\end{equation}
Using equation~(\ref{eq:r1}) and after some rearranging of terms, we
reach the final result for the limiting case
\begin{equation}
    \lim_{S\rightarrow \infty}\frac{\sigma_u}{u}=\left(\frac{\sigma_B}{B}\right)
    \left(\frac{B}{S}\right) (1-u) \left[1-\left(1-\frac{1}{u}\right)
      \cos i\cos\theta\right]\;.
\end{equation}
This expression demonstrates that, even in the case of perfect prior
knowledge of the geometry of the system and of an infinite number of
accumulated photons, the accuracy in the fractional measurement of the
neutron star compactness is proportional to the uncertainty in our
measurement of the background ($\sigma_{\rm B}/B$) and to the ratio of
the background-to-source counts ($B/S$).

\subsection{Utilizing Additional Constraints}

In the previous section, we considered the case of minimal information
that can be obtained from the pulse profiles, assuming isotropic
emission, slow rotation, and no distance or mass information for the
pulsar. We now discuss the additional constraints that can be achieved
on the neutron star radius and compactness, if any of these
assumptions is relaxed.

We first explore the more favorable case when the emission is
anisotropic, i.e., $h\ne 0$. This is also a more realistic
consideration because of the strong temperature gradients in the
bombarded atmospheres of the polar cap regions. The uncertainties
discussed in the previous section all improve because of the presence
of one additional constraint related to the measurement of the second
harmonic of the profile. The two independently measurable quantities
then become
\begin{eqnarray}
\frac{V_1}{Q}&=&\frac{(1+2hq)v}{q+h(q^2+v^2/2)}\nonumber\\
\frac{V_2}{Q}&=&\frac{hv^2}{2\left[q+h(q^2+v^2/2)\right]}\;.
\label{eq:ratios}
\end{eqnarray}
These two independent pieces of information allow for a unique
determination of the auxiliary quantities $q$ and $v$, which can be
used to determine the compactness $u$ and the combination of the
angles $i$ and $\theta$ (since these angles enter $q$ and $v$ only as
a product of their sines and cosines, see eq.[19]). The fractional
uncertainties $\sigma_Q/Q$ and $\sigma_V/V$, then, will give rise to
similar uncertainties in the two ratios in equation~(\ref{eq:ratios}).
Individually, these translate into uncertainties in the compactness
$u$ in much the same way as in the previous section, following
Equations~(22)-(26). However, combining them allows for smaller
uncertainties on the compactness $u$ and the combination of the angles
$i$ and $\theta$. More harmonics can be measured also in the case 
of a moderately spinning star where Doppler effects are detectable 
(e.g., Psaltis \& \"Ozel 2014), leading to similar tightening of 
the uncertainties. 

A similar situation arises when the polar-cap size $dS$ and the
measured intensity of radiation $I_o$ are measured spectroscopically
and the distance to the source $D$ is known a priori. This is
achievable for some of the representative NICER targets where a
precise measurement of the distance is possible via pulsar timing
parallax or very long baseline interferometry (e.g., \ofour, see
Deller et al.\ 2008; Verbiest et al.\ 2008). In this case, instead of
taking the ratios of the various amplitudes, it becomes possible to
directly invert equations~(\ref{eq:general}) to obtain the auxiliary
quantities $u$ and $v$. As before, these additional constraints lead
to much tighter constraints on the compactness $u$ and the combination
of the angles $i$ and $\theta$.  The independent knowledge of one of
these angles then completely solves the system of equations.

\begin{figure}[t]
\begin{center}
  \psfig{file=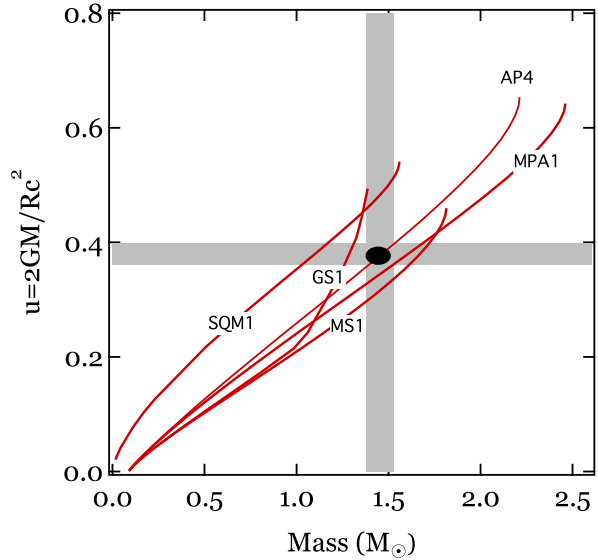,width=3.5in,clip=}
\end{center}
  \caption{The solid curves show the neutron-star compactness as a
    function of its mass, for a number of representative equations of
    state. The vertical shaded band represents a dynamical measurement
    of the mass of a $1.45 M_\odot$ neutron star with a 5\%
    uncertainty. The horizontal shaded band represents a measurement
    of its compactness via pulse-profile modeling, with a 5\%
    uncertainty. Achieving such an accuracy is sufficient to
    distinguish between different equations of state.}
\label{fig:eos}
\mbox{}
\end{figure}

Finally, we consider in Figure~\ref{fig:eos} the case where the mass
of the pulsar is known through other means, such as a dynamical
measurement of the pulsar mass in a binary system. This is indeed a
likely situation, as a large fraction of millisecond pulsars are in
binary systems with white dwarf companions. Currently, there are 17
millisecond pulsars with precisely measured masses, including some of
the NICER targets, and this number is growing rapidly (see a recent
compilation in \"Ozel \& Freire 2016). This figure shows that
combining the measurement of the compactness $u$ with a prior
knowledge of the mass with an uncertainty that is comparable to that
of \ofour\ ($M=1.44(7)\;M_\odot$; Reardon et al.\ 2016) is sufficient
to distinguish between different equations of state.

\section{Conclusions}

In this paper, we investigated the effect of instrument or sky
background on the determination of neutron star compactness and radius
from a measurement of the shapes of pulse profiles from
rotation-powered pulsars. Specifically, we made use of analytic
formulae for the pulse shapes of moderately spinning neutron stars in
the Schwarzschild metric to connect the observables from the profile
to the properties of the neutron star. We showed that, in the case of
isotropic emission and not utilizing any distance or mass information
for the pulsar, the ratio of harmonics can be used to constrain the
neutron star compactness. Furthermore, we showed that knowledge of the
background at a few percent level for a background-to-source countrate
ratio of 0.2 and a total accumulation of $10^6$ counts allow for a
measurement of the neutron star compactness to better than $10\%$
uncertainty for most of the parameter space. These uncertainties
become even smaller for more favorable geometries where the harmonic
amplitudes are higher, as is the case for prototypical NICER targets
such as \ofour. For these geometries, a $\lesssim 6\%$ uncertainty in
the compactness is achievable even when the background-to-signal ratio
becomes comparable to $B/S \approx 2$ (this is, e.g., a factor of two
worse than what is expected for \oo).

Polar caps of rotation powered pulsars emit thermally in the soft
X-rays because they are heated from above by particle bombardment from
the magnetosphere. In this case, it is expected from model atmosphere
calculations that the emission from the surface will be beamed. This
anisotropy allows for more fractional harmonic amplitudes to be
measured compared to the isotropic case and can reduce the uncertainty
in the compactness even further for the same requirements on the
background. Similarly, utilizing distance information to the pulsar,
when it is known precisely, and when the instrument's effective area
is sufficiently well calibrated, allows for a measurement of the
absolute, rather than fractional, amplitudes, increasing the number of
directly and independently determined observables to two or three,
depending on the beaming of emission. Finally, a mass measurement for
the pulsar obtained through pulsar timing enables a precise
measurement of the neutron star radius when combined with the
compactness measured from the pulse profile. All of these will lead to
a more accurate determination of the dense matter equation of state
through means that are different from the earlier spectroscopic
measurements (see \"Ozel et al. 2015).

\end{document}